\documentclass[journal=jpccck,manuscript=article]{achemso}
\usepackage{times}
\usepackage{epsfig}
\usepackage{graphicx}
\usepackage{psfrag}
\usepackage{ae}
\usepackage{amsmath,amssymb}
\usepackage[usenames]{color}
\usepackage{float}

\title{ Thermodynamic and dynamic anomalous behavior in the  
TIP4P/$\varepsilon$ water model}

\author{ Ra\'ul Fuentes-Azcatl}
\author{Marcia C. Barbosa}

\affiliation{Instituto de F\'{\i}sica, Universidade Federal do Rio Grande do Sul, \\
Caixa Postal 15051, 91501-970, Porto Alegre, RS, Brazil}
\begin{document}
\begin{abstract}

The model   Tip4p/$\varepsilon$  for water  is 
tested for the presence of thermodynamic and dynamic 
anomalies. Molecular dynamic simulations for this model
were performed and we show that for
this system the density versus temperature at
constant pressure exhibits a maximum.
In
addition  we also show
that the  diffusion coefficient versus density at constant temperature  has
a maximum and a minimum. The anomalous behavior of the density
and of the diffusion coefficient obey the water hierachy.  The
results for the Tip4p-$\epsilon$  are consistent with experiments and 
when compared with the Tip4p-2005 model show similar results a variety
of physical properties and better performance for the dielectric constant.

\end{abstract}

\section{Introduction}

Water is a fascinating molecule. Even though present
in our everyday life, it shows a  number
of  properties that are still not well 
described~\cite{Eisenberg, Chaplin}.
For example, most liquids contract upon cooling. This is not the case of water, a 
liquid where the specific volume at ambient pressure starts to increase
 when cooled below $4^C$ at atmospheric pressure~\cite{Ke75}. 
In addition, in a certain
range of pressures, water also exhibits an anomalous increase of 
compressibility 
and of the specific heat upon cooling~\cite{Pr87,Pr88,Ha84}.
Water also has dynamic anomalies. Experiments  show that the diffusion 
constant, $D$, increases on compression at low temperature, $T$,  up to a 
maximum $D_{\rm max} (T)$ at $p =p_{D\mathrm{max}}(T)$. The behavior 
of normal liquids, with $D$  decreasing on 
compression, is 
restored in water only at high $p$, e.g. for 
$p > p_{D\mathrm{max}}\approx 1.1$ kbar  at $10^o$C ~\cite{An76}.

In addition to the measured anomalies
of water, theoretical analysis predicted anomalies~\cite{Chaplin,Po92} that are located
in regions of the pressure versus temperature phase diagram of difficult
access experimentally. Consequently, simulations became an interesting tool 
to test these theories.
Then the  challenge faced when developing a computational strategy
is to design a model that would be general enough 
to describe the different behaviors of water and simple enough
to be  computationally treatable.
The later prerequisite at the moment preclude the consideration of 
quantum effects
and polarization. Both polarization and quantum effects, however,  seems to play  a relevant
role in the anomalous properties of water particularly when
charges and interfaces are present.
In order to circumvent
this difficulty without loosing 
the simplicity required for simulation
purposes, a number of 
atomistic models has been developed with the assumption that 
 polarization and quantum
effects were included in an averaged way.

These atomistic models are characterized
by representing
the charges in water
by two, three, four or even five points.  Then, the 
interactions were modelled by 
a classical Lennard-Jones for 
the hardcore interactions and the
electrostatic interactions for the charges. This
leads to the following parameters that need
to be specified: the values and positions
of the charges, the positions and masses of the atoms and
the energy and size for the LJ interaction.  Then, the crucial step
in the modelling process
is the choice of the set of  quantities used to fit these 
parameters. This set
should be small but appropriated to guarantee that the model 
reproduces
as many properties of water as possible at least in 
a certain range of temperatures and pressures.

Within the non-polarizable models the  4-site 
form represented an advance. It was first proposed by Bernal and Fowler~\cite{Be33} 
along with a
set of parameters based on calculations for properties
of the monomer, dimer, and ice. The fours
points are the position of the oxygen and hydrogens and 
the location , $M$, of the negative charge. Within
the Tip4p each hydrogen carries a positive charge, $q_H$,
while the negative charge, $q_M$ is located at a position
$r_{OM}$ from the oxygen between the two hydrogens. The
angle between the oxygens and the hydrogens, $104.52^o$,
and the distance between the oxygen and the hydrogen, $r_{OH}=0.9572$, where
fixed to reproduce the ice structure.  
The Bernal and Fowler model, however, gives a poor results for 
the liquid properties at room $25^oC$ and atmospheric
pressure. A reparametrization
of this model gave rise to the TIP4P~\cite{Jo83} model that shows 
good agreement  with the density at $25^oC$ and $1\;atm$
and an excellent value for the vaporization enthalpy.
In addition this model provides a 
reasonable description of some solid phases
and reproduces qualitatively the phase 
diagram~\cite{guillot02,Ab05,vega09,vega11} while the results
for the SPC/E and TIP5P models are quite poor.
Unfortunately it gives 
a value that is too low for the temperature
of maximum density and of melting.
Then, it became clear that good model of water should provide
the behavior of the liquid, particularly the
value of the  density anomaly at
atmospheric pressure, and a
reasonable description of the solid phases. For that purpose
the TIP4P/2005~\cite{Ab05} was created. It was designed to match
the density at the temperature of maximum density but 
yields a slightly low melting
temperature and a somewhat large vaporization enthalpy.

In order to test the TIP4P/2005 model
against other options,  Vega et al.~\cite{vega11} have
compared a number of the non-polarizable models.
The strategy was to select a set of water properties
and compared the results obtained 
by different models with the experiments.
They established that the best 
model to reproduce the properties  they 
have selected 
 is the four sites TIP4P/2005~\cite{Ab05,vega09} followed by the three 
sites SPC/E model~\cite{spce}. The only 
drawback of these
models is that they do not give a good description 
for the dielectric constant of water. Since the dielectric
constant is fundamental for understanding the behavior of 
mixtures of water and other substances, particularly polar molecules, these
water models are not appropriated to analyse these mixtures.

In order to circumvent this
difficulty without loosing the 
advantages of the  TIP4P/2005~\cite{Ab05,vega09}, Fuentes 
et al.~\cite{tip4pe}
 developed the non-polarizable 
TIP4P/$\epsilon$ rigid model. This
potential is parametrized to give 
the experimental value of the  density and of the dielectric 
constant at $4^oC$ and atmospheric pressure. This new model 
showed that it is  
capable of  reproducing some
thermodynamic quantities~\cite{tip4pe}
obtained by the  TIP4P/2005~\cite{Ab05,vega09}. In
addition it  
gives a good agreement with the experiments for   the isothermal
 compressibility and  dielectric constant at different 
pressures and temperatures what is not observed in the
non-polarizable models.

In addition to the thermodynamic 
anomalies, water also show 
a singular mobility.  
While experimental results show that 
the diffusion coefficient of water decreases 
with decreasing pressures up to crystallization, simulations
with SPC/E water
show that this system if kept liquid reaches 
a minimum~\cite{Er01, Ne01,Ne02} at negative pressures.
Then the pressure and the temperature of the maximum and the minimum
of the mobility define a region of diffusion anomaly. This
region englobes  the density anomaly defining
the hierarchy of the anomalies~\cite{Er01, Ne01,Ne02}. 
This hierarchy has been employed to conceptualize
the mechanism behind the thermodynamic and dynamic
unusual behavior of water.
This result suggests that 
the thermodynamic and the dynamic anomalies
are not independent, but are related by the
competition of two length scales: bonding and 
non bonding~\cite{Ol06}.

Therefore it would be  desirable that a model
for water would be capable to capture not only the thermodynamic
anomalies but also the dynamical anomalous behavior of water.
In this paper we test if the
TIP4P/$\epsilon$ model
also shows the dynamic anomalous
region in the pressure versus temperature
phase diagram observed in water. We compute the
diffusion coefficient, $D$, versus temperature for 
various densities and temperatures. Then the location 
in the pressure versus temperature 
of the maximum and minimum of 
the diffusion coefficient $D$ are compared with 
the density extrema and checked if the TIP4P/$\epsilon$ has
the hierarchy observed in the experiments. Our results
are also compared with the 
TIP4P/2005~\cite{Ab05,vega09, vega11} model.
Finally a summary of the thermodynamic, dynamic and 
structural properties of this model is compared with experiments
and with the TIP4P/2005 mode in the spirit of the {\it grading}
proposed by Vega et al.~\cite{vega11}.

The remaining of this paper goes as follows. In the section 2 the
force fields for the TIP4P/2005 and TIP4P/$\epsilon$ are presented. 
In the section 3 the simulations are explained and in section 4 
results are analysed. Conclusions are shown in section 5.

\section{The Models}
\label{sec:Models}

\begin{figure}[!h]
\centering
\includegraphics[scale=0.35]{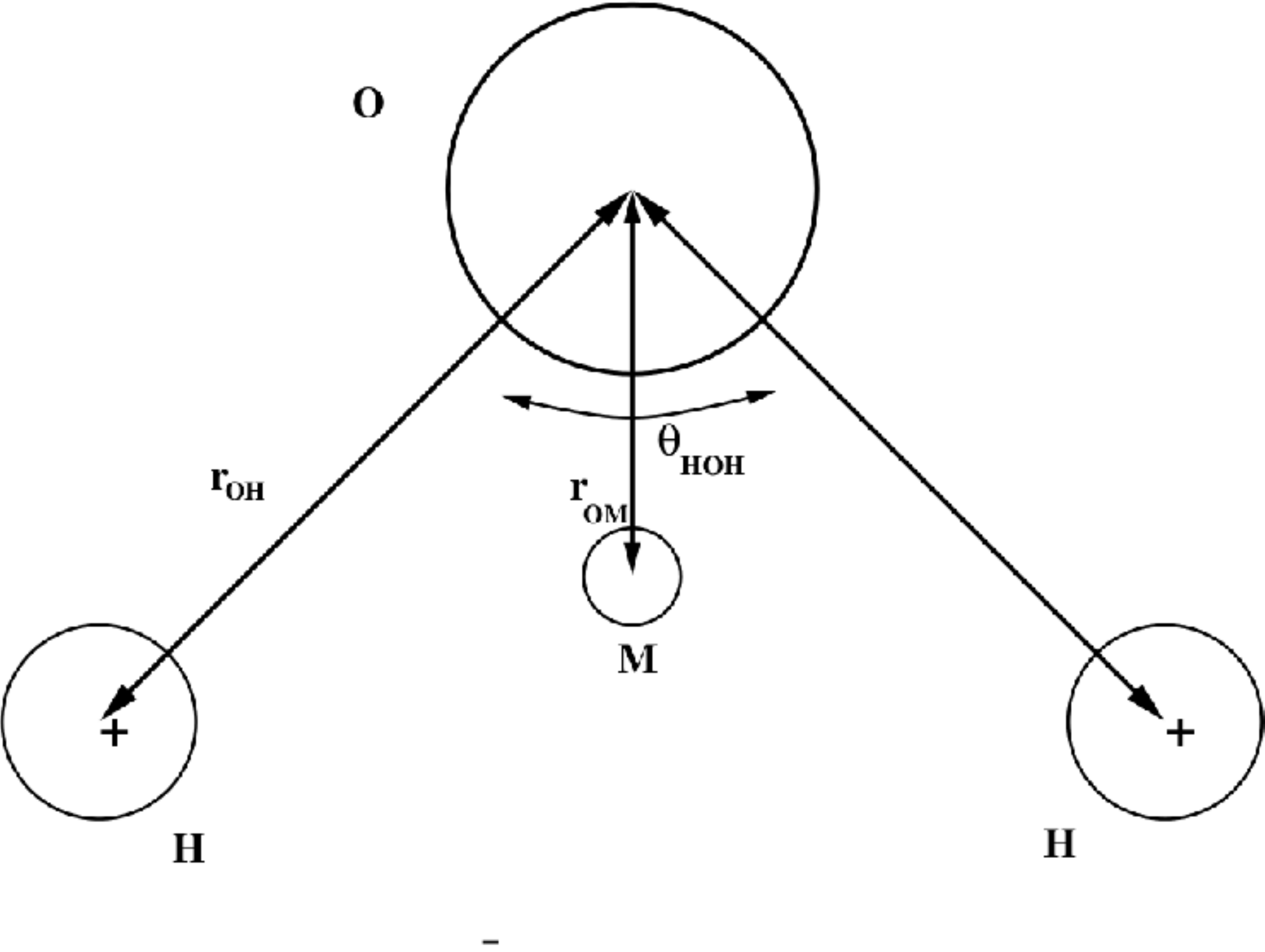}
\caption{Model of four points for water: one oxygen, two
hydrogens and a point M where the negative
charge is located.  }
\label{fig:model}
\end{figure}
The different propositions
for the four site models have in common the
format illustrated in the Figure~\ref{fig:model}. 
The system is represented by the oxygen and the two hydrogens. 
The distance between the oxygen and
the hydrogen is given by  $r_{OH}=0.9572$ while
the angle between the two hydrogens and the oxygen is
$\theta_{HOH}=104.52^o$. These quantities were set in order
to give the appropriated ice form. The oxygen has mass $m_O=15.999\ g/mol$ and
the hydrogens have mass $m_O=1.008\ g/mol$. The shared electrons between
the hydrogens and the oxygen are closer to the oxygen. This is represented by 
the two positive
charges, $q_H$, one at each  hydrogen and a negative 
charge $q_M\approx 2q_H$ at a point M distant $r_{OM}$ from the
oxygen and located between the two hydrogens. 
Each water
molecule $i$ has a kinetic energy, $K_i$ given by 
\begin{equation}
\label{kin}
K_{i} = \frac{1}{2} m_O v_i^2
\end{equation}
where $v_i$ is the velocity of the molecule. 
Two water molecule interact through a potential
with two contributions, a Lennard-Jones (LJ) between the oxygens
and electrostatic interactions between the hydrogens and 
the negative charge at the point M, namely
\begin{equation}
\label{ff}
u(r_{ij}) = 4\epsilon_{OO} 
\left[\left(\frac {\sigma_{OO}}{r_{ij}}\right)^{12}-\left 
(\frac{\sigma_{OO}}{r_{ij}}\right)^6\right] + \frac{1}{4\pi\epsilon_0}
\sum_{i=1}^3\sum_{j=1}^3\frac{q_i q_j}{r_{ij}}
\end{equation}
\noindent where $r_{ij}$ is the distance between atom  $i$ and 
$j$, $q_i$ is the electric charge of atom $i$, $\epsilon_0$ is 
the permittivity of vacuum,  $\epsilon_{OO}$ is the LJ energy scale
 of the oxygen 
interactions  and  $\sigma_{OO}$ the diameter for an $OO$ pair. The 
model has one LJ site and charge on the oxygen atom and additionally 
a charge on every hydrogen atom.  

In this paper some thermodynamic, dynamic and structural properties of 
two force fields are compared: TIP4P/2005 and the TIP4P/$\epsilon$.
For the first model the parameter, $r_{OM}$, $\epsilon_{OO}$,
$q_H$, ($q_M = 2q_H$) and $\sigma_{OO}$ are selected by 
imposing that the model reproduces the maximum density 
at $T=4^oC$ and atmospheric pressure. For the TIP4P/$\epsilon$ 
the parameters are chosen so the model not only reproduces
the density but also  the dielectric contant at $T=4^oC$ 
at atmospheric pressure.
The parameters of these two force fields are given in 
the Table \ref{tip4pe}.

\begin{table}
\caption{Force field parameters for water models. The charge in site $M$ is $q_M=-(2 q_H)$.  }
\label{tip4pe}
\begin{tabular}{|cccccc|}
\hline\hline
 Model & $q_H/e$  & $q_M/e=2q_H/e$& $r_{OM}/\AA$ & $\sigma_{OO}$/\AA & $(\epsilon_{OO}/k_B)$/K\\
\hline
TIP4P/$\varepsilon$ & 0.9572 & 1.054& 0.105 & 2.4345 & 93\\
TIP4P/2005 & 0.9572 & 1.1128& 0.1546 & 2.305 & 93.2\\
\hline
\end{tabular}
\end{table}

\section{Simulation details}
\label{sec:Simulation}

All the simulations in this work have been done for
a system of 500 molecules and employing molecular
dynamic simultions in the NVT ensemble with the  package 
GROMACS 4.5.\cite{gromacs}. The equations of motion are solved 
using the leap-frog 
algorithm\cite{allen,gromacs} and the time step used was 2 fs. The Lennard-Jones  
potential has been switched from 10 $\AA$ up to a cut-off 
distance of 10 $\AA$. Long 
range corrections were applied to the Lennard-Jones part of 
the potential (for 
both the energy and pressure).

Ewald summations were used to deal with electrostatic contributions. The 
real 
part of the Coulombic potential is truncated at 
10 $\AA$. The Fourier component 
of the Ewald sums was evaluated by using the particle mesh Ewald (PME) 
method~\cite{PME} 
using a grid spacing of 1.2 $\AA$ and a fourth degree polynomial for the 
interpolation. The simulation box is cubic throughout the whole 
simulation and the 
geometry of the water molecules kept constant using the shake 
procedure~\cite{shake}. Temperature has been set to the desired 
value with a Nos\'e Hoover thermostat~\cite{nhc}.

The diffusion coefficient is
calculated using the mean-square displacement averaged over different
initial times,namely
\begin{equation}
\label{eq:Deltar2}
\langle \Delta r(t)^{2} \rangle = \langle [r(t_0+t)-r(t_0)]^2\rangle.
\end{equation}
 
From Eq. (\ref{eq:Deltar2}), the diffusion coefficient may be obtained 
as follows:
\begin{equation}
D=\lim_{t\to\infty}\langle \Delta r(t)^{2} \rangle/6t.
\end{equation}

The static dielectric constant is computed from the fluctuations\cite{neumann} of the total dipole moment {\bf M},
\begin{equation}
\epsilon=1+\frac{4\pi}{3k_BTV} (<{\bf M}^2>-<{\bf M}>^2)
\end{equation}
\noindent where  $k_B$ is the Boltzmann constant and $T$ the absolute temperature. The dielectric constant is obtained for long simulations at constant density and temperature or at constant temperature and pressure. 
 
\section{Results}
\label{sec:Results}

\begin{figure}[!h]
\centering
\includegraphics[scale=0.65]{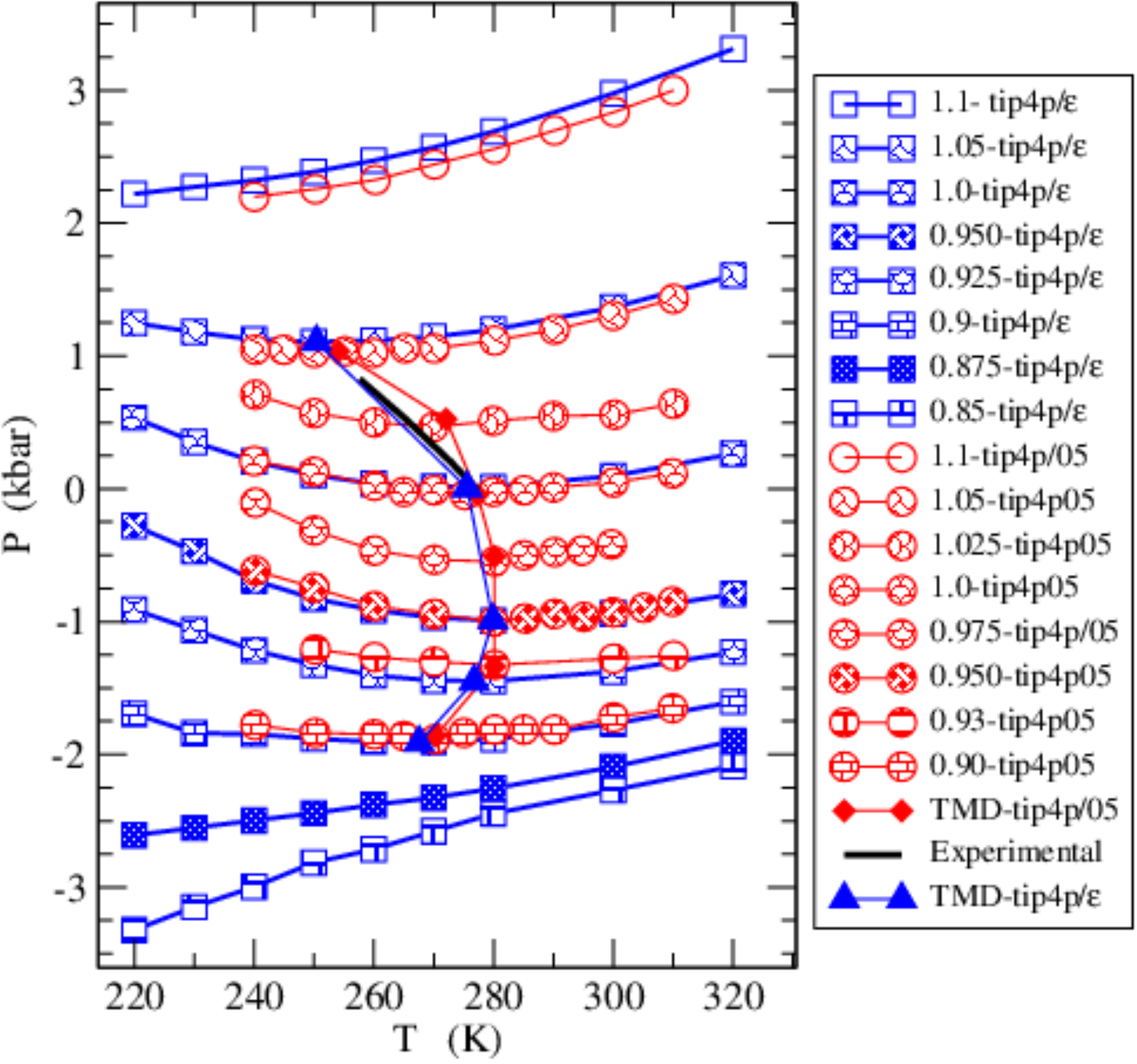}
\caption{Pressure versus temperature phase diagram. The red 
circles are the isochores for the TIP4P/2005 model while
the blue squares are the isochores for the TIP4P/$\epsilon$ model.
The minimum of the isochores show the TMD location for each model.
The black line indicates the experimental results for the TMD.
indicate isochores\cite{Fine}.}
\label{fig:pt-density}
\end{figure}
\begin{figure}[!h]
\centering
\includegraphics[scale=0.55]{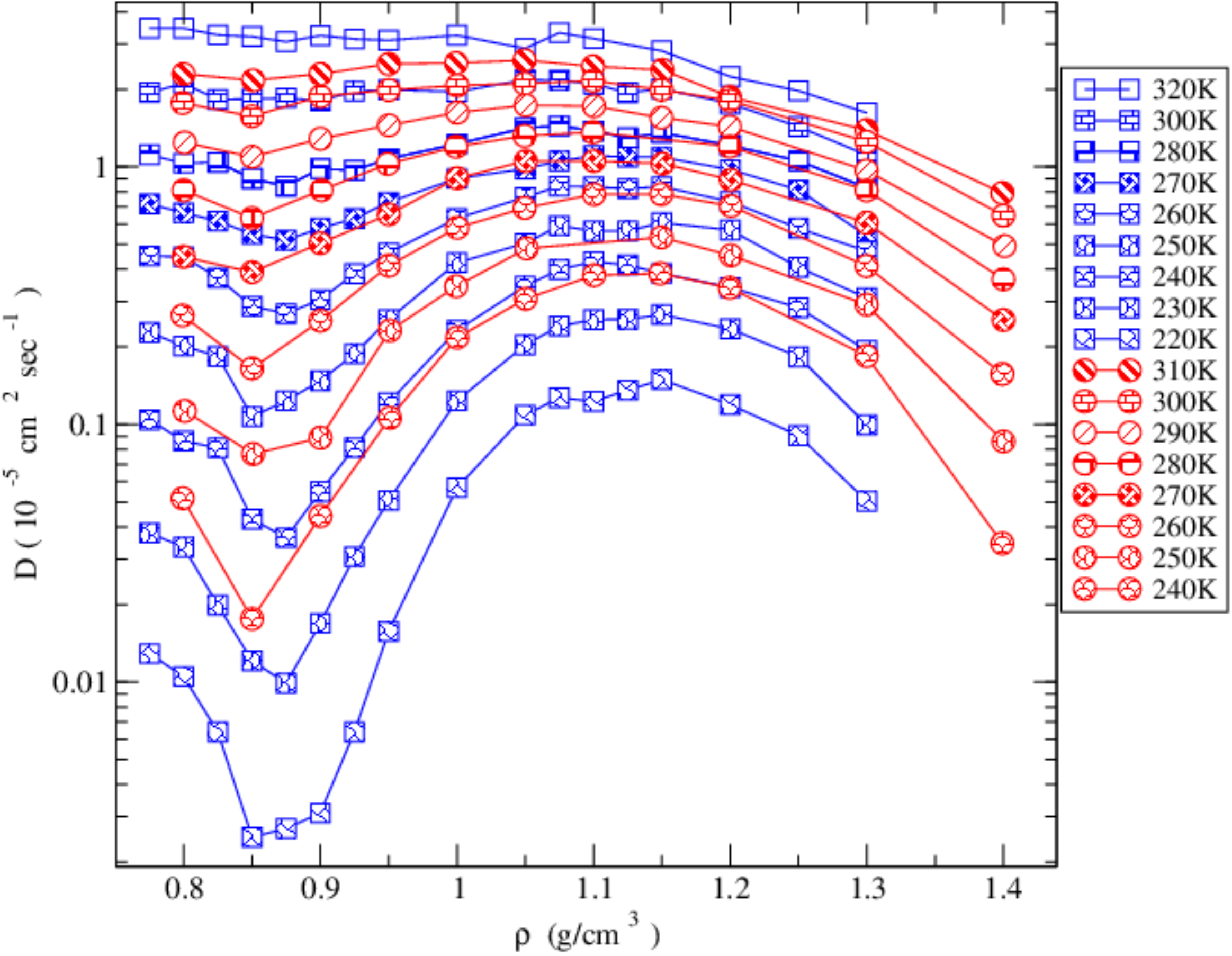}
\caption{Diffusion coefficient versus density for 
various temperatures. The red circles illustrated $D$ for the TIP4P/2005 model while the blue circles show the TMD for the  TIP4P/$\epsilon$ model.
Both maximum of $D$.}
\label{fig:diffusion}
\end{figure}

\begin{figure}[!h]
\centering
\includegraphics[scale=0.75]{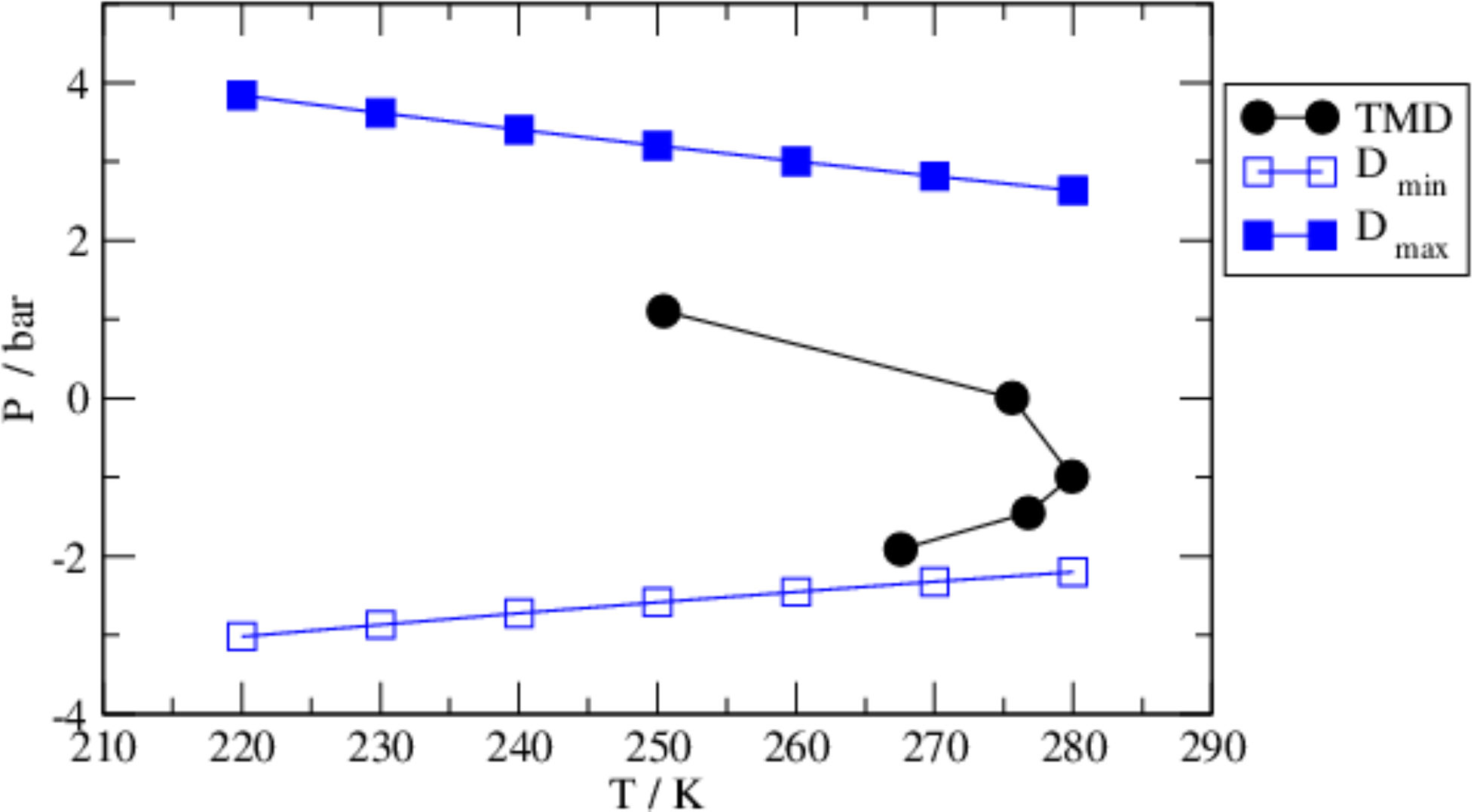}
\caption{Pressure versus temperature phase diagram illustrating
the temperature of maximum density (black filled circles)  and the maximum (blue filled squares)  and minimum (blue empty squares) of 
the diffusion coefficient for the  TIP4P/$\epsilon$ model.}
\label{Fig3}
\end{figure}

\begin{figure}[!h]
\centering
\includegraphics[scale=0.65]{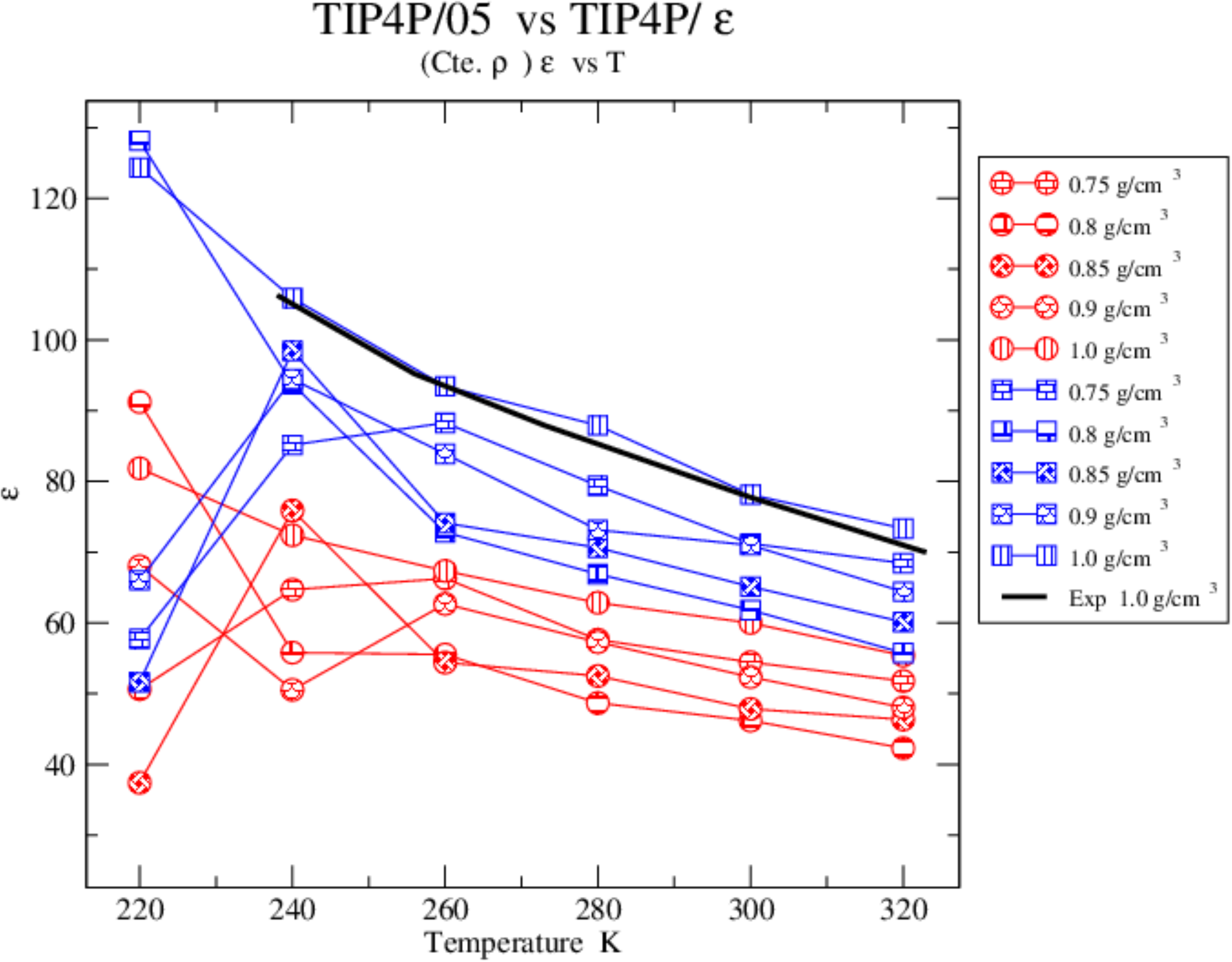}
\caption{Dielectric constant  versus temperature for
different densities from $0.75g/cm^3$ to $1g/cm^3$
(from bottom to top) for the TIP4P/2005 (red circles)
and for the  TIP4P/$\epsilon$ (blue squares) models.  The
experimental\cite{Prini} results (black solid line) are shown for 
$1g/cm^3$.   }
\label{fig:die-T}
\end{figure}

 \begin{figure}[!htb]
\centering
\includegraphics[scale=0.65]{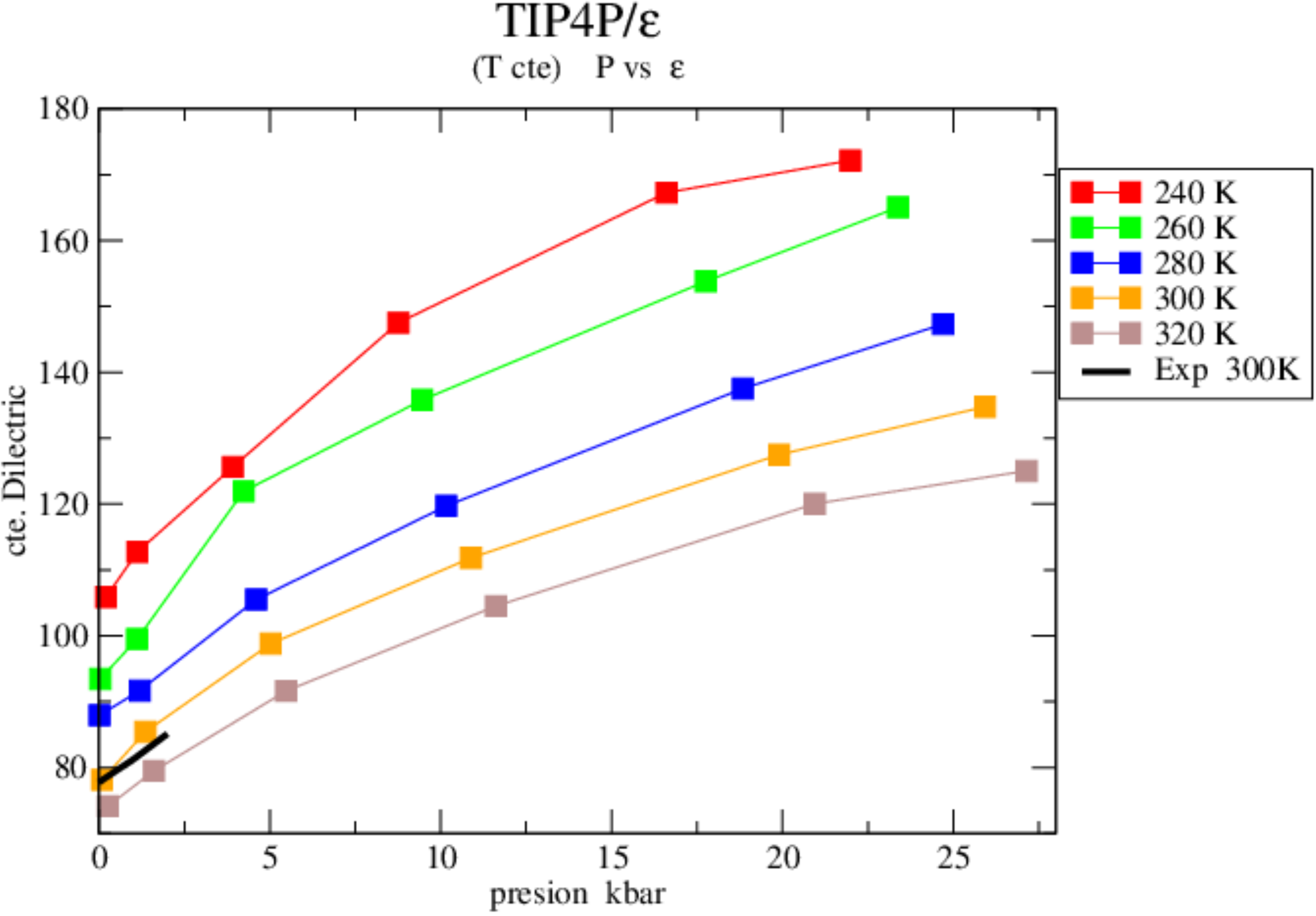}
\caption{Dielectric constant versus pressure for 
$T=240, 260, 280, 300, 320 \; K$
for the TIP4P/$\epsilon$ model (circles) and for $T=300\;K$
for experiments\cite{Prini} (black squares). }
\label{fig:die-P}
\end{figure}

First, the temperatures of maximum density for the
different pressures were computed for
both TIP4P/2005 and TIP4P/$\epsilon$  models
as follows. In the NVT ensemble this is
done by relating the minimum of the isochores 
at the pressure versus temperature
phase diagram.  Using the Maxwell relation,
\begin{eqnarray}
(\frac{\partial V}{\partial T})_P &= & - (\frac{\partial P}{\partial T})_V
(\frac{\partial V}{\partial P})_T \nonumber \\
(\frac{\partial \rho}{\partial T})_P &= &- (\frac{\partial P}{\partial T})_V(\frac{\partial \rho}{\partial P})_T
\label{eq:Maxwell}
\end{eqnarray}
the maximum of $\rho(T)$ versus
temperature at constant pressure given
by $(\partial \rho/\partial T)_P=0$ is
equivalent to the minimum of  $(\partial P/\partial T)_{\rho}=0$. While
 the former is suitable for NPT-constant
experiments/simulations the latter is more convenient for our
NVT-ensemble study, thus adopted in this work.

  Figure~\ref{fig:pt-density} illustrates the pressure 
versus temperature phase digram where the isochores 
for the TIP4P/2005 and for te TIP4P/$\epsilon$  models
are shown as red circles and blue squares respectively.
The minimum of the isochores are also illustrated as 
red losangles for the  TIP4P/2005 model and blue triangles for
the TIP4P/$\epsilon$ model. These lines
locate the Temperature of Maximum Density (TMD).
The simulations give a good agreement
with the experimental results for the TMD 
represented by a black solid 
line. The 
two models are quite equivalent for the location
of the TMD what is not surprising since both
are adjusted to give the location of the density
maximum  at
atmospherica pressure.

In addition to the thermodynamic anomaly the diffusion anomaly
is also analyzed. Figure~\ref{fig:diffusion} shows the 
diffusion coefficient versus 
density for both models for a range of temperatures.
For both models de diffusion coefficient versus
density graph has a maximum
and minimum for various temperatures.
The values of the temperature of
the maximum and minimum diffusion coefficients
are consistent with the values 
obtained by experiments~\cite{An76}. The pressures
for the maximum and minimum $D$, however, give higher values 
when compared
with the experimental results~\cite{An76} what might 
be attributed to the rigidity of both models. 

The hierarchy of anomalies of the TIP4P/$\epsilon$ model shown
in the Figure~\ref{fig:p-t-anomalies} confirms the 
predicted behavior that the region in the pressure 
versus temperature phase diagram in which the 
diffusion anomaly is present involves the region 
where the density anomaly appears~\cite{Er01}.

We also check the behavior of the dielectric constant, $\epsilon$,
with the temperature and density. Figure~\ref{fig:die-T}
shows $\epsilon$ as a function of the 
temperature for different densities for the TIP4P/2005
(red circles)  and TIP4P/$\epsilon$ (blue squares)  models.
The TIP4P/2005 model shows much lower values of dielectric
constant when compared with the TIP4P/$\epsilon$  model
and experiments~\cite{tip4pe}. 
The fact that the TIP4P/$\epsilon$ gives a good estimate
for the dielectric constant at room pressure and 
temperature is not surprising since the model was
fitted to give this result. It is interesting to
observe that for this result is also preserved for 
other values of temperatures.

Since the TIP4P/$\epsilon$ does not provide a good evaluation
of the pressure for the maximum of the diffusion anomaly
it is important to verify if at high pressures the 
dielectric constant fails to agree with the experiments.
Figure~\cite{fig:die-P} illustrates the dielectric constant versus 
pressure for different temperatures for the TIP4P/$\epsilon$ model
and experiments for $T=300\;K$. The agreement between simulations
and experiments is good for the low pressures. Experimental
data for higher pressures still need to be further explored.

In recent years Vega et al.~\cite{vega11} proposed to evaluate the
performance of water models by 
a measure. The models
received a grade from zero to ten by
checking  how a finite group of  properties from the liquid, solid and
gas phases predicted
by the model agree with experimental
results. In addition to  equilibrium
thermodynamic properties, the measure includes dynamic properties and
phase transition predictions. 
In order to answer to the logical criticism that the TIP4P/$\epsilon$ model
performes well in computing the dielectric constant 
but might fail  in other properties in which TIP4P/2005
gives good agreements with experiments~\cite{vega15}, the measured proposed
by  Vega  et al.~\cite{vega11} was computed for the TIP4P/$\epsilon$ model.

Table~\ref{vega-parameters} shows
the performance of the TIP4P/$\epsilon$ model compared with
the performance of the  TIP4P/2005 for the properties 
proposed by  Vega et al.~\cite{vega11}.
Our results indicate that TIP4P/$\epsilon$ gives
good results not only for the dielectric constant, density and diffusion 
anomalies but also for the selected properties illustrated in the table.

\begin{table}
\caption{}
\scalebox{0.8}[0.65]{
\begin{tabular}{|l|c|c|c|c|c|c|}
\hline
	&	Experiment&	Quantity	&Quantity	&		&	Score	&	Score	\\
Property	&	data	&	TIP4P/05	&	TIP4P/$\epsilon$	&	$\%$ Tol. 	&	TIP4P/05		&TIP4P/$\epsilon$	\\
\hline													
\multicolumn{7}{|l|}{Enthalpy of phase change / kcal mol$^{-1}$}\\													
\hline													
$\Delta$H$_{melt}$	&	1.44	&	1.16	&	1.24	&	5	&	6	&	7	\\
$\Delta$H$_{vap}$	&	10.52	&	11.99	&	11.74	&	2.5	&	4	&	5	\\
\hline													
\multicolumn{7}{|l|}{Critical point properties}\\													
\hline													
T$_C$/K	&	647.1	&	640	&	675.45	&	2.5	&	10	&	8	\\
$\rho_C$/g cm${^{-3}}$	&	0.322	&	0.337	&	0.2993	&	2.5	&	8	&	7	\\
p$_C$/bar	&	220.64	&	146	&	136	&	5	&	3	&	2	\\
\hline													
\multicolumn{7}{|l|}{Surface tension/mN m$^{-1}$}\\													
\hline													
$\gamma_{300K}$	&	71.73	&	69.3	&	69	&	2.5	&	9	&	8	\\
$\gamma_{450K}$	&	42.88	&	41.8	&	43.8	&	2.5	&	9	&	9	\\
\hline													
	\multicolumn{7}{|l|}{Melting properties}\\												
\hline													
T$_m$/K	&	273.15	&	252	&	240	&	2.5	&	7	&	5	\\
$\rho_{liq}$/g cm$^{-3}$	&	0.999	&	0.993	&	0.994	&	0.5	&	9	&	9	\\
$\rho_{sol}$/g cm$^{-3}$	&	0.917	&	0.921	&	0.920	&	0.5	&	9	&	9	\\
dp/dT (bar K$^{-1}$)	&	-137	&	-135	&	-134	&	5	&	10	&	10	\\
\hline													
	\multicolumn{7}{|l|}{Orthobaric densities and temperature of maximun density \textbf{TMD}}\\												
\hline													
\textbf{TMD}/K	&	277	&	278	&	277	&	2.5	&	10	&	10	\\
$\rho_{298K}$/g cm$^{-3}$	&	0.997	&	0.993	&	0.99628	&	0.5	&	9	&	10	\\
$\rho_{400K}$/g cm$^{-3}$	&	0.9375	&	0.93	&	0.9368	&	0.5	&	8	&	10	\\
$\rho_{450K}$/g cm$^{-3}$	&	0.8903	&	0.879	&	0.8842	&	0.5	&	7	&	9	\\
\hline													
	\multicolumn{7}{|l|}{Isothermal compressibility / 10$^-6$ bar$^{-1}$)}\\												
\hline													
$\kappa_T$ [1 bar; 298 K]	&	45.3	&	46	&	45.8	&	5	&	10	&	10	\\
$\kappa_T$ [1 bar;360 K]	&	47	&	50.9	&	49.1	&	5	&	8	&	9	\\
\hline													
	\multicolumn{7}{|l|}{Gas properties}\\												
\hline													
$\rho_{v}$[350 K] (bar)	&	0.417	&	0.13	&	0.026	&	5	&	0	&	0	\\
$\rho_{v}$[450 K] (bar)	&	9.32	&	4.46	&	2.64	&	5	&	0	&	0	\\
B2[450 K] (cm$^3$ mol$^{-1}$ )	&	-238	&	-476	&	-438	&	5	&	0	&	0	\\
\hline													
	\multicolumn{7}{|l|}{Heat capacity at constant pressure/cal mol$^{-1}$K$^{-1}$}\\												
\hline													
C$_p$[liq 298 K; 1 bar]	&	18	&	21.1	&	19.1	&	5	&	7	&	9	\\
C$_p$[ice 250 K; 1 bar]	&	8.3	&	14	&	11.9	&	5	&	0	&	1	\\
\hline													
	\multicolumn{7}{|l|}{Static dielectric constant}\\												
\hline													
$\varepsilon$[liq; 298 K]	&	78.5	&	58	&	78.3	&	5	&	5	&	10	\\
$\varepsilon$[I$_h$; 240 K]	&	107	&	53	&	63	&	5	&	0	&	2	\\
Ratio	&	1.36	&	0.91	&	0.80	&	5	&	3	&	2	\\
\hline													
	\multicolumn{7}{|l|}{T$_m$-TMD-T$_c$. ratios}\\												
\hline													
T$_m$[I$_h$]/T$_c$	&	0.422	&	0.394	&	0.355	&	5	&	9	&	7	\\
TMD/T$_c$	&	0.428	&	0.434	&	0.410	&	5	&	10	&	9	\\
TMD-T$_m$(K)	&	3.85	&	26	&	37	&	5	&	6	&	5	\\
\hline													
	\multicolumn{7}{|l|}{Densities of ice polymorphs/g cm$^{-3}$}\\												
\hline													
$\rho$[I$_h$ 250 K; 1 bar]	&	0.92	&	0.921	&	0.919	&	0.5	&	10	&	10	\\
$\rho$[II 123 K; 1 bar]	&	1.19	&	1.199	&	1.196	&	0.5	&	8	&	9	\\
$\rho$[V 223 K; 5.3 kbar]	&	1.283	&	1.272	&	1.275	&	0.5	&	8	&	9	\\
$\rho$[VI 225 K; 11 kbar]	&	1.373	&	1.38	&	1.377	&	0.5	&	9	&	9	\\
\hline													
	\multicolumn{7}{|l|}{EOS high pressure}\\												
\hline													
$\rho$[373 K; 10 kbar]	&	1.201	&	1.204	&	1.202	&	0.5	&	10	&	10	\\
$\rho$[373 K; 20 kbar]	&	1.322	&	1.321	&	1.318	&	0.5	&	10	&	9	\\
\hline													
	\multicolumn{7}{|l|}{Self-diffusion coefficient/cm$^2$s$^{-1}$ }\\												
\hline													
ln D$_{278K}$	&	-11.24	&	-11.27	&	-11.3657	&	0.5	&	9	&	8	\\
ln D$_{298K}$	&	-10.68	&	-10.79	&	-10.77	&	0.5	&	8	&	8	\\
ln D$_{318K}$	&	-10.24	&	-10.39	&	-10.3	&	0.5	&	7	&	9	\\
E$_a$ k$_J$ mol$^{ -1}$	&	18.4	&	16.2	&	16.3	&	5	&	8	&	8	\\
\hline													
	\multicolumn{7}{|l|}{Shear viscosity / mPa s}\\												
\hline													
$\eta$[1 bar; 298 K]	&	0.896	&	0.855	&	0.863	&	5	&	9	&	9	\\
$\eta$[1 bar; 373 K]	&	0.284	&	0.289	&	0.307	&	5	&	10	&	8	\\
\hline													
	\multicolumn{7}{|l|}{Orientational relaxation time / ps}\\												
\hline													
$\tau_2^{HH}$ [1 bar; 298 K]	&	2.36	&	2.3	&	2.31	&	5	&	9	&	10	\\
\hline
	\multicolumn{7}{|l|}{Structure}\\											\hline	
$\chi^2$(F(Q))	&	0	&	8.5	&	8.5	&	5	&	8	&	8	\\
$\chi^2$(overall)	&	0	&	14.8	&	14.8	&	5	&	7	&	7	\\
\hline													
\multicolumn{5}{|l|}{Phase diagram}					&	8	&	8	\\
\hline													
\hline													
\multicolumn{3}{|l|}{Overall score (out of 10)}					&		&		&	7.13	&	7.31	\\
\hline

	\end{tabular}}
\label{table-calif3}
\end{table}

\section{Conclusions}

In this paper we have computed explicitly the behavior of 
the density and diffusion anomalies for the 
TIP4P/$\epsilon$ model showing that it gives 
similar results when compared with the TIP4P/2005 model.

Then, the behavior of the dielectric constant with
the temperature and pressure was analyzed and compared
with experimental results showing better agreement 
than the values obtained by other non-polarizable models.

Finally, the set of properties proposed by Vega el al. was
computed. 
Our results that the performance of the TIP4P/$\epsilon$
model is similar to the performance of the  TIP4P/2005 model~\cite{vega11}
with the exception of the dielectric constant for which the 
TIP4P/$\epsilon$ shows a better agreement with the experimental
results. We hope that this model might be suitable for 
studying mixtures and confined systems where the 
dielectric constant play an important role.

\section{Acknowledgment}

We thank the Brazilian agencies CNPq, INCT-FCx, and Capes for the financial support. We also  thank the Centro Latino-Americano de F\'isica (CLAF), Secretar\'ia de Ciencia, Tecnolog\'ia e Innovaci\'on del Distrito Federal de M\'exico (SECITIDF) and CONACYT (Project No. 232450) for financial support .

\section{References}
\bibliographystyle{elsarticle-num}
\bibliography{<your-bib-database>}


\end{document}